\RequirePackage{ifpdf}
\ifpdf 
\documentclass[pdftex]{sigma}
\else
\documentclass{sigma}
\fi

\begin{document}
\allowdisplaybreaks

\renewcommand{\PaperNumber}{066}

\FirstPageHeading

\ShortArticleName{Quantum Entanglement and Projective Ring Geometry}

\ArticleName{Quantum Entanglement\\ and Projective Ring Geometry}

\Author{Michel PLANAT~$^\dag$, Metod SANIGA~$^\ddag$ and Maurice R. KIBLER~$^{\S}$}
\AuthorNameForHeading{M. Planat, M. Saniga and M.R. Kibler}

\Address{$^\dag$~Institut FEMTO-ST, CNRS/Universit\'e de Franche-Comt\'{e}, D\' epartement LPMO,\\ 
$\phantom{^\dag}$~32 Avenue de l'Observatoire, F-25044 Besan\c con Cedex, France} 
\EmailD{\href{mailto:planat@lpmo.edu}{planat@lpmo.edu}} 

\Address{$^\ddag$~Astronomical Institute, Slovak Academy of Sciences,\\
$\phantom{^\ddag}$~SK-05960 Tatransk\' a Lomnica, Slovak Republic}
\EmailD{\href{mailto:msaniga@astro.sk}{msaniga@astro.sk}}

\Address{$^{\S}$~Institut de Physique Nucl\'{e}aire de Lyon, IN2P3-CNRS/Universit\'{e} Claude Bernard Lyon 1,\\ 
$\phantom{^\S}$~43 Boulevard du 11 Novembre 1918, F-69622 Villeurbanne Cedex, France}
\EmailD{\href{mailto:kibler@ipnl.in2p3.fr}{kibler@ipnl.in2p3.fr}}

\ArticleDates{Received June 13, 2006, in f\/inal form August 16,
2006; Published online August 17, 2006}

\Abstract{The paper explores the basic geometrical properties of
the observables characterizing two-qubit systems by employing a
novel projective ring geometric approach. After introducing the
basic facts about quantum complementarity and maximal quantum
entanglement in such systems, we demonstrate that the $15 \times 15$
multiplication table of the associated four-dimensional matrices
exhibits a so-far-unnoticed geometrical structure that can be
regarded as three pencils of lines in the projective plane of
order two.   In one of the pencils, which we call the kernel, the
observables on two lines share a base of Bell
states.  In the
complement of the kernel, the eight vertices/observables are
joined by twelve lines which form the edges of a cube.  A
substantial part of the paper is devoted to showing that the
nature of this geometry has much to do with the structure of the
projective lines def\/ined over the rings that are the direct
product of $n$ copies of the Galois f\/ield
$GF(2)$, with $n = 2, 3$ and 4.}


\Classification{81P15; 51C05; 13M05; 13A15; 51N15; 81R05} 

\section{Introduction}

``Seriousness'' of quantum theory for addressing the most
fundamental aspects of reality has invariably been at the
forefront of theoretical explorations of most prominent scholars
\cite{Einstein,Bohm,Bell,Kochen,Peres,Mermin}, being f\/irmly established by
experiment in 1982 \cite{Aspect}. Two measurements described by
non-commuting observables are inherently uncertain and this led
Einstein, Podolsky and Rosen~\cite{Einstein} to question the
completeness of quantum theory versus the reality of  both
observed physical quantities. Using counterfactual arguments
applied to distant experimental set-ups they introduced (and
immediately rejected) the notion of underlying wholeness, which
shortly after gave rise to the concept of quantum entanglement
\cite{Schroedinger}. Bohr believed that no serious conclusion can
be drawn from the comparison of thought experiments dealing with
mutually incompatible (i.e., non-commuting) observables and thus
practically ignored the paradox, proposing another
view/paradigm--quantum complementarity \cite{Bohr}. Since the
work of Bohm \cite{Bohm} and Bell \cite{Bell}, the ``puzzles''
of quantum theory have mainly been discussed within a discrete
va\-riab\-le setting of spin-$\frac{1}{2}$ particles. In essence,
Bell's theorems \cite{Mermin} imply that either the recursive
(counterfactual) reasoning about possible experiments should be
abandoned, or non-contextual assumptions (implicit in the EPR
locality arguments) are to be challenged, or both. One of the
simplest illustrations of quantum ``mysteries'', which also
provides a very economical proof of the Bell--Kochen--Specker
theorem \cite{Bell,Kochen}, employs a $3 \times 3$ array of
nine observables characterizing two spin-$\frac{1}{2}$ particles
\cite{Mermin}. The three operators in any row or column of such a
square, commonly referred to as the Mermin  ``magic'' square,
are mutually commuting, allowing the recursive reasoning to be
used, but the algebraic structure of observables contradicts that
of their eigenvalues~\cite{Mermin}. This contradiction stems from
the following two basic features of the structure of the square:
complementarity between the observables located in two distinct
rows and two of the columns and the maximal entanglement of the
observables in one of the columns.

The basic facts about quantum complementarity and maximal quantum
entanglement for two spin-$\frac{1}{2}$ particles (or two-qubits,
using the language of quantum information theory) are given in
Section~\ref{compl}. In Section~\ref{Fano} we demonstrate that the
$15 \times 15$ multiplication table of the associated
four-dimensional (generalized Pauli spin) matrices exhibits a
so-far-unnoticed geometrical structure, which can be regarded as
three pencils of lines in the projective plane of order two (the
Fano plane) \cite{Polster}. These three pencil-conf\/igurations,
each featuring seven points/observables, share a~line (called the
reference line), and any line comprises three observables, each
being the product of the other two, up to a factor $-1$, $i$ or
$-i$ ($i^2=-1$). All the three lines in each pencil carry mutually
commuting operators; in one of the pencils, which we call the
kernel, the observables on two lines share a base of maximally
entangled states. The three operators on any line in each pencil
represent a row or column of some of  Mermin's  ``magic''
squares, thus revealing an inherent geometrical nature of the
latter \cite{SanigaPlanat1,Aravind}. In the complement of
the kernel, the eight vertices/observables are joined by twelve
lines which form the edges of a cube. The lines between the kernel
and the cube are pairwise complementary, which means that each
vertex/observable is linked with six other ones.

Some of these intriguing geometrical features can be recovered, as
shown in detail in Section~\ref{sec4}, in terms of the structure of the
projective line def\/ined over the f\/inite ring $GF(2)^{\otimes n}$,
with $n = 2,3,4$, $GF(2)\cong \mathcal{Z}_2$ denoting the Galois
f\/ield with two elements and $\otimes n$ representing the direct
product of $n$ such f\/ields. After recalling some basics on the
concept of a projective ring line and the associated concepts of
neighbour and distant, we illustrate its basic properties over the
ring $GF(2)^{\otimes 2}$ and show that the corresponding line
reproduces nicely all the basic qualitative properties of a Mermin
square (Section~\ref{twiceGF}). In order to account for a more intricate
geometrical structure of the kernel and the cube, one has to
employ the lines corresponding to $n=3$ (Section~\ref{threetimesGF}) and $n=4$
(Section~\ref{fourtimesGF}), respectively. Although these two lines provide us
with important insights into the structure of the two operator
conf\/igurations, it is obvious we will have to look for a higher
order ring line in order to get a more complete geometrical
picture of two-qubit systems.

\section[Quantum complementarity, maximal entanglement and mutually unbiased bases]{Quantum 
complementarity, maximal entanglement\\ and mutually unbiased bases}
\label{compl}

\begin{table}[t]\small
\caption{Multiplication between the elements of $\mathcal{A}$.}

\vspace{-3mm}
\begin{center}
\begin{tabular}{||l|rrrr|rrrr||}
\hline\hline
 $*$& $0$ & $1$ & $2$ & $3$ & $6$ & $14$ & $9$ & $12$  \\
\hline
0& $0$ & $1$ & $2$ & $3$ & $6$ & $14$ & $9$ & $12$  \\
1& $1$ & $0$ & $3$ & $2$ & $-i 14$ & $i 6$ & $-i 12$ & $i 9$  \\
2& $2$ & $3$ & $0$ & $1$ & $i 9$ & $i 12$ & $-i 6$ & $-i 14$  \\
3& $3$ & $2$ & $1$ & $0$ & $12$ & $-9$ & $-14$ & $6$  \\
\hline
6& $6$ & $i14$ & $-i9$ & $12$ & $0$ & $-i1$ & $i2$ & $3$  \\
14& $14$ & $-i 6$ & $-i 12$ & $-9$ & $i 1$ & $0$ & $-3$ & $-i 2$  \\
9& $9$ & $i12$ & $i6$ & $-14$ & $-i2$ & $-3$ & $0$ & $-i1$  \\
12& $12$ & $-i9$ & $i14$ & $6$ & $3$ & $i2$ & $i1$ & $0$  \\
\hline\hline
\end{tabular}
\label{table1}
\end{center}

\vspace{-3mm}

\caption{Multiplication between the elements of $\mathcal{B}$.}

\vspace{-3mm}
\begin{center}
\begin{tabular}{||l|rrrr|rrrr||}
\hline\hline
 $*$& $4$ & $7$ & $11$ & $13$ & $5$ & $10$ & $15$ & $8$  \\
\hline
4& $0$ & $1$ & $i2$ & $i3$ & $6$ & $14$ & $-i9$ & $-i 12$  \\
7& $1$ & $0$ & $i3$ & $i2$ & $-i 14$ & $i6$ & $ -12$ & $9$  \\
11& $-i2$ & $-i3$ & $0$ & $1$ & $9$ & $12$ & $i 6$ & $ i14$  \\
13& $-i 3$ & $-i 2$ & $1$ & $0$ & $-i 12$ & $i 9$ & $14$ & $-6$  \\
\hline
5& $6$ & $i14$ & $9$ & $i 12$ & $0$ & $-i 1$ & $2$ & $-i 3$  \\
10& $14$ & $-i 6$ & $12$ & $-i9$ & $i 1$ & $0$ & $i 3$ & $2$  \\
15& $i9$ & $-12$ & $-i6$ & $14$ & $2$ & $-i3$ & $0$ & $-i1$  \\
8& $i12$ & $9$ & $-i14$ & $-6$ & $i3$ & $2$ & $i1$ & $0$  \\
\hline\hline
\end{tabular}
\label{table2}
\end{center}
\vspace{-3mm}

\caption{Multiplication between the elements of $\mathcal{A}$ and $\mathcal{B}$.}

\vspace{-3mm}
\begin{center}
\begin{tabular}{||l|rrrr|rrrr||}
\hline\hline
 $*$& $0$ & $1$ & $2$ & $3$ & $6$ & $14$ & $9$ & $12$  \\
\hline
4& $4$ & $7$ & $-i11$ & $-i13$ & $5$ & $10$ & $i15$ & $i8$  \\
7& $7$ & $4$ & $-i13$ & $-i11$ & $-i 10$ & $i5$ & $ 8$ & $-15$  \\
11& $11$ & $13$ & $i4$ & $i7$ & $-i15$ & $-i8$ & $5$ & $10$  \\
13& $13$ & $11$ & $i7$ & $i4$ & $-8$ & $15$ & $-i10$ & $i5$  \\
\hline
5& $5$ & $i10$ & $15$ & $i 8$ & $4$ & $-i7$ & $11$ & $-i13$  \\
10& $10$ & $-i5$ & $8$ & $-i15$ & $i7$ & $4$ & $i13$ & $11$  \\
15& $15$ & $i8$ & $5$ & $i10$ & $i11$ & $13$ & $-i4$ & $-7$  \\
8& $8$ & $-i15$ & $10$ & $-i5$ & $-13$ & $-i11$ & $7$ & $-i4$  \\
\hline\hline
\end{tabular}
\label{table3}
\end{center}\vspace{-5mm}
\end{table}

\looseness=1
Bohr's concept of quantum complementarity \cite{Bohr} has recently
received great attention in relation with the problem of f\/inding
complete sets of so-called mutually unbiased bases (MUBs). Two
observables are complementary if precise knowledge of one of them
implies that all possib\-le outcomes of measuring the other are
equally probable. The eigenstates of such observables are
non-orthogonal quantum states and in an attempt to distinguish
between them any gain of information is only possible at the
expense of introducing disturbances~-- a property of crucial
importance in quantum cryptography. Let $O$ be an observable in a
f\/inite dimensional Hilbert space of dimension $q$, represented by
a Hermitian matrix with multiplicity-free eigenvalues such that
its eigenvectors $|b\rangle$  belong to an orthonormal basis $B$.
Let $O'$ be a prepared complementary observable with eigenvectors
$|b'\rangle$ in a basis $B'$. If $O$ is measured, the probability
to f\/ind the system in the state $|b \rangle$  is $|\langle
b|b'\rangle|^2=1/q$. If the latter relation holds for any two
pairs $|b\rangle$ and~$|b'\rangle$, then the two bases are said to
be mutually unbiased. It can be shown that in order to fully
recover the density matrix of a set of copies of an unknown
quantum state, we need at least $q+1$ measurements performed on
complementary observables. This number also represents the upper
bound for the cardinality of distinct MUBs to exist, and such
(complete) sets have so far been constructed only for $q=p^m$,
with $p$ being a prime number and $m$ a positive integer, the most
elegant techniques employed being those using additive characters
over Galois f\/ields $GF(p^m)$ (for $p > 2$) and Galois rings
$R=GR(4^m)$ (for $p=2$) \cite{Planat05}. This property was in
\cite{Saniga04} postulated to be equivalent to a long standing
combinatorial problem of the non-existence of projective planes of
orders dif\/fering from powers of primes, a work that can be
regarded as one of the f\/irst implementations of f\/inite algebraic
geometrical objects/structures into the context of quantum bits
(qubits). A closely related SU(2) ``polar'' recipe for
constructing MUBs has recently been proposed
\cite{KiblerPlanat06}. It is also worth mentioning that MUBs are a
key ingredient in numerous attempts of accounting for entanglement
related ``paradoxes'' \cite{SanigaPlanat1,Aravind}.

\begin{figure}[t]
\centerline{\includegraphics[width=5cm]{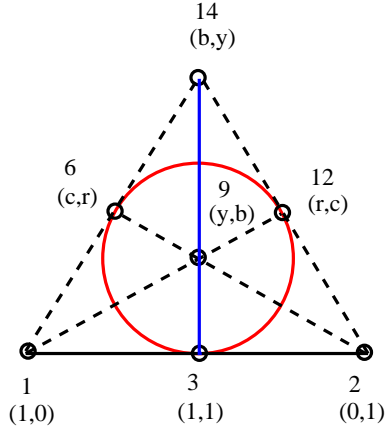}}
\caption{The ``kernel of entanglement'' for two-qubit systems as a
pencil of lines (one represented by the circle) in the Fano plane.
The points of the conf\/iguration correspond to the nontrivial
observables of equation~(\ref{set A}). The extra labelling refers to
the points of the projective line of $GF(2)^{\otimes 3}$ (see
Section~\ref{threetimesGF}).} \label{figure1}
\end{figure}

\begin{figure}[t]
\centerline{\includegraphics[width=5cm]{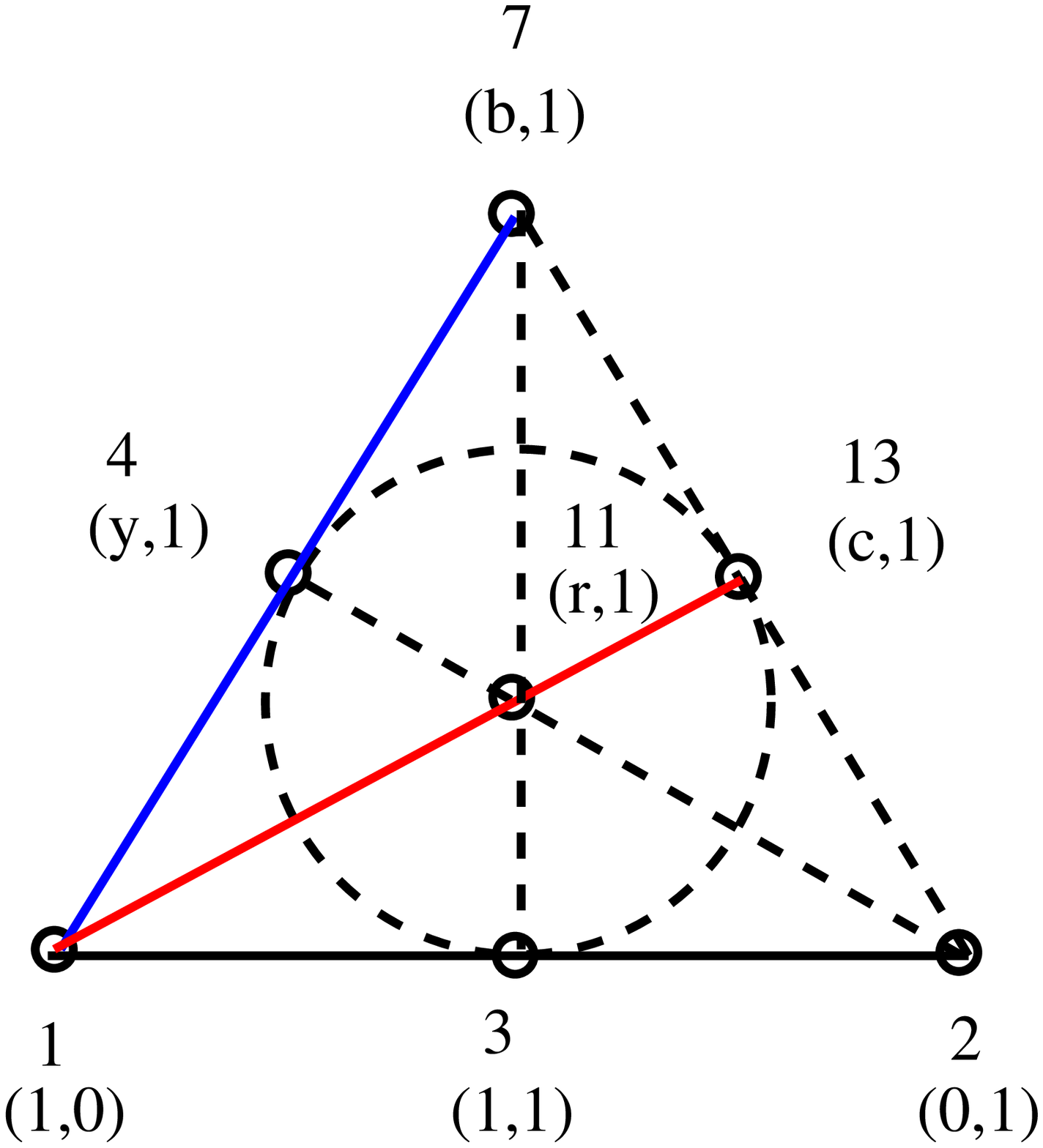}\qquad
\includegraphics[width=5cm]{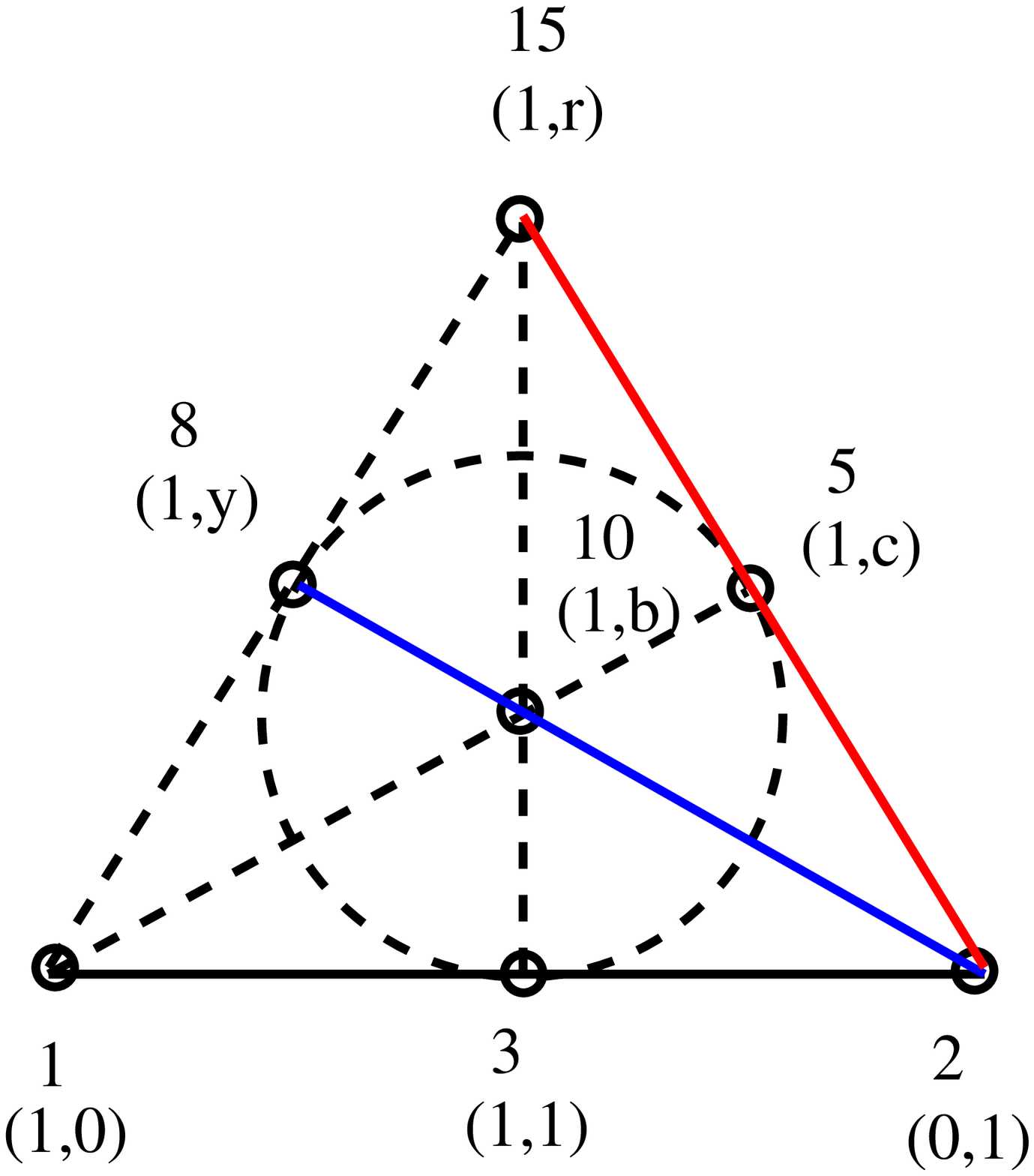}}
\caption{Two pencils of lines reproducing the product rules of the observables given Table~\ref{table3}. 
The points, apart from 1 to 3, are now the observables from (\ref{set B}). 
The extra labelling refers again to the points of the projective line over $GF(2)^{\otimes 3}$ 
(see Section~\ref{threetimesGF}).}
\label{figure2}
\end{figure}

There exists an equivalence between the unbiasedness of sets of
bases and particular sets of mutually commuting operators sharing
a base of eigenvectors. Let us consider the partitioning of the
$15$ observables attached to two spin-$\frac{1}{2}$ particles into
the following $5$ $(=4+1)$ mutually unbiased sets arranged in rows \cite{Lawrence}
\begin{gather}
 \begin{array}{ccc}
1_2\otimes \sigma_z & \sigma_z\otimes 1_2 & \sigma_z \otimes \sigma_z\\
\sigma_x\otimes 1_2  &1_2\otimes \sigma_y & \sigma_x \otimes \sigma_y\\
\sigma_x \otimes \sigma_z & \sigma_z \otimes \sigma_x & \sigma_y \otimes \sigma_y\\
1_2 \otimes \sigma_x & \sigma_y \otimes 1_2 & \sigma_y \otimes \sigma_x\\
\sigma_y \otimes \sigma_z & \sigma_x \otimes \sigma_x & \sigma_z \otimes \sigma_y\\
\end{array}
=
\begin{array}{ccc}
1&2&3\\
4&5&6\\
7&8&9\\
10&11&12\\
13&14&15\\
\end{array}
\label{MUBs}
\end{gather}
where $1_2$ is the $2 \times 2$ unit matrix and $\sigma_x$,
$\sigma_y$ and $\sigma_z$ are the classical Pauli matrices. All
the observables in (\ref{MUBs}) have doubly-degenerate
eigenvalues, $\pm 1$, and each row gives rise to an orthogonal
base; the bases represented by the 3rd and 5th rows are entangled.
Every operator in a row is the product of the other two, i.e.\
$1*2=3$, $4*5=6,\ldots$ (here $*$ stands for the matrix product),
but no similar rule seems to exist between operators in two
dif\/ferent rows. The $3 \times 3$ arrays of observables of Mermin's
type, mentioned in the introduction, are of the following forms
\begin{gather}
\begin{array}{ccccccccccccccccccc}
1&2&3&\mbox{}&\mbox{}&1&2&3&\mbox{}&\mbox{}&1&4&7&\mbox{}&\mbox{}&1&11&13&\\
4&10&14&\mbox{}&\mbox{}&13&15&14&\mbox{}&\mbox{}&2&5&15&\mbox{}&\mbox{}&2&10&8&\\
7&8&9&\mbox{}&\mbox{}&11&5&9&\mbox{}&\mbox{}&3&6&12&\mbox{}&\mbox{}&3&12&6&\\
\end{array}
\label{Mermin}
\end{gather}
In each of the above arrays, the observables in every row or
column are mutually commuting, each being the product of the other
two {\it except for} the last column where a minus sign appears,
i.e., $3*14=-9$; the product of the three operators in each row
and the f\/irst two columns thus yields $+1_2$, whereas for the
third column it is $-1_2$. In view of our subsequent
considerations it is useful to enumerate the orthogonal bases
attached to rows and columns of the f\/irst Mermin square on the
left-hand side of equation~(\ref{Mermin}), omitting, for the sake of
simplicity, a normalization factor and denoting the sign of
eigenvalues of the corresponding operators by subscripts:
\begin{gather}
\begin{array}{ccccc}
[1,2,3]:&(1,0,0,0)_{+++}&(0,1,0,0)_{-+-}&(0,0,1,0)_{+--}&(0,0,0,1)_{--+}  \\ \nonumber
=&|00\rangle & |01\rangle & |10\rangle &  |11\rangle \\ \nonumber
[4,10,14]: &(1,1,1,1)_{+++}&(1,-1,1,-1)_{+--} &(1,1,-1,-1)_{-+-}&(1,-1,-1,1)_{--+}  \\ \nonumber
[7,8,9]:&(1,1,1,-1)_{+++}&(1,-1,1,1)_{+--} &(1,1,-1,1)_{-+-}&(1,-1,-1,-1)_{--+}  \\ \nonumber
[1,4,7]:&(1,0,1,0)_{+++}&(1,0,-1,0)_{+--} &(0,1,0,1)_{-+-}&(0,1,0,-1)_{--+}  \\ \nonumber
[2,10,8]:&(1,1,0,0)_{+++}&(1,-1,0,0)_{+--} &(0,0,1,1)_{-+-}&(0,0,1,-1)_{--+}  \\ \nonumber
[3,14,9]:&(1,0,0,1)_{++-}&(1,0,0,-1)_{+-+} &(0,1,1,0)_{-++}&(0,-1,1,0)_{---}  \\ \nonumber
=&|00\rangle +|11\rangle & |00\rangle-|11\rangle  & |01\rangle+|10\rangle  &  |01\rangle -|10\rangle \\ \nonumber
\end{array}\!
\label{Eigenvectors}
\end{gather}
For the f\/irst two arrays of (\ref{Mermin}), an orthogonal base of the Bell states is associated with the third column;
for the other two squares, the Bell states are carried by the operators in the third row as follows
\begin{gather}
\begin{array}{ccccc}
[3,6,12]:&(1,0,0,i)_{+++}&(1,0,0,-i)_{+--} &(0,1,i,0)_{-++}&(0,1,-i,0)_{---}  \\ \nonumber
=&|00\rangle +i|11\rangle & |00\rangle-i|11\rangle  & |01\rangle+i|10\rangle  &  |01\rangle -i|10\rangle \\ \nonumber
\end{array}
\label{Eigenvectors2}
\end{gather}

\section[Algebra and geometry of two spin-$\frac{1}{2}$ particles]{Algebra and geometry 
of two spin-$\boldsymbol{\frac{1}{2}}$ particles}
\label{Fano}

\looseness=1
The four dif\/ferent representations of Mermin's square,
equation~(\ref{Mermin}), give us important hints about the existence
of an underlying algebraic geometrical principle governing
interaction of two spin-$\frac{1}{2}$ particles. The base line
$(1,2,3)$ is common to all the four arrays, each of the remaining
operators from $4$ to $15$ appears twice, and the (Bell) entangled
triples $(3,14,9)$ and $(3,6,12)$ form a column and a row,
respectively. Our goal in this section is to reveal this hidden
geometry.

\begin{figure}[t]
\centerline{\includegraphics[width=6cm]{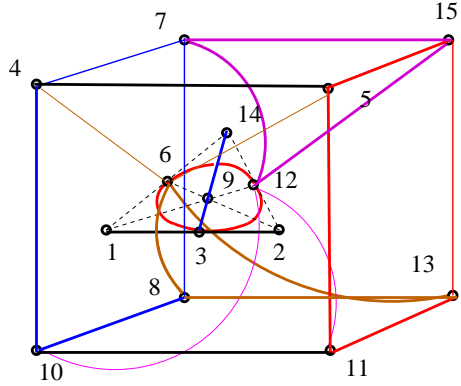}}
\caption{A geometry of a system of  two spin-$\frac{1}{2}$
particles. The ``inner'' observables are arranged as shown in
Fig.~\ref{figure1}, whereas the ``outer'' ones form a cubic conf\/iguration.
A couple of ``inner'' observables commute with eight ``outer''
ones forming two ``complementary'' four-tuples. They exist
three such pairs: $(6,12)$, $(9,14)$ and $(1,2)$. 
The ``inner/outer'' relation is, however, illustrated for the f\/irst pair
only. The concerned triples of observables 
which are entangled are represented with thicker lines. } \label{figure3}
\end{figure}

To this end in view, we f\/irst add to the set given by equation~(\ref{MUBs}) the identity
operator $0\equiv 1_2 \otimes 1_2$, obtaining
\begin{gather}
\mathcal{S}=\{0,1,2,\ldots, 15\}, \label{set 1/2}
\end{gather}
and partition this set into two subsets $\mathcal{A}$ and $\mathcal{B}$, where
\begin{gather}
\mathcal{A}=\{0,1,2,3,6,9,12,14\}
\label{set A}
\end{gather}
comprises the ``computational'' operators $\mathcal{C}=\{0,1,2,3\}$ and the ``entangled'' operators $\mathcal{E}=\{6,9,12,14\}$, and
\begin{gather}
\mathcal{B}=\{4,5,7,8,10,11,13,15\},
\label{set B}
\end{gather}
which can also be partitioned into two subsets of cardinality four
as shown in Table~\ref{table2}. Next we create the multiplication tables for
the elements of $\mathcal{A}$ (Table \ref{table1}), $\mathcal{B}$ (Table~\ref{table2})
and those of both sets (Table~\ref{table3}) in order to see that the
following properties hold for two observables $O$ and~$O'$
\begin{gather*}
O,O' \in \mathcal{A} ~~{\rm or}~~ O,O' \in \mathcal{B} ~~\Rightarrow~~ O*O' \in \mathcal{A}, \nonumber \\
O \in \mathcal{A}~~{\rm and}~~O' \in \mathcal{B}~~\Rightarrow~~
O*O' \in \mathcal{B}. 
\end{gather*}
\looseness=-1 One immediately recognizes that the multiplication table
of $\mathcal{C}$ is, except for a factor $-1$ or $\pm i$,
isomorphic to the addition table of the Galois f\/ield
$GF(4)\cong GF(2)[x]/\langle x^2+x+1 \rangle$ and that of $\mathcal{A}$
to the addition table of the Galois f\/ield $GF(8) \cong GF(2)[x]/\langle
x^3+x+1 \rangle$. The set $GF(8)^* \equiv GF(8)\setminus\{0\}$ is
a cyclic group generated by a single element and its
representation in terms of $3$-tuples in $GF(2)^{\otimes 3}$
provides the coordinates of seven points of the projective plane
of order two, the Fano plane \cite{PlanatBielefeld}. Hence, we can
identify the elements of the set~$\mathcal{A}$, omitting the
trivial one~(0), with the points of such a plane and obtain the
conf\/iguration shown in Fig.~\ref{figure1}. In this f\/igure, three operators
are on a line if and only if the product of two of them equals,
again apart from a factor $-1$ or $\pm i$, the remaining one (see
Table~\ref{table1}), with the understanding that full/broken lines join
commuting/non-commuting operators. The three full lines,
distinguished from each other by dif\/ferent colours, form a
so-called pencil, i.e., the set of lines passing through the same
point, called the base point -- see, e.g., \cite{Polster}, and,
in this particular case, each of them is endowed with the triple
of operators carrying {\it Bell} states. The essence of quantum
entanglement between two spin-$\frac{1}{2}$ particles is thus
embodied in a very simple geometry! If, furthermore, the 
set~$\mathcal{B}$ and Table~\ref{table3} are taken into account, we get two
analogous conf\/igurations, as depicted in Fig.~\ref{figure2}.
However, these conf\/igurations dif\/fer crucially from the f\/irst one
as any full line in either of them contains triples of operators
that share an {\it un}entangled orthogonal base of eigenvectors.
All in all, the f\/ifteen operators 1 to 15 are thus found to form a
remarkable conf\/iguration comprising the seven elements of the
kernel of Fig.~\ref{figure1} and the ten elements of 
the ``outer shell'' forming a~cube, as shown in Fig.~\ref{figure3}.

\section[Entanglement and finite ring geometry]{Entanglement and f\/inite ring geometry}\label{sec4}

\subsection[The $GF(2)^{\otimes 2}$ geometry of the Mermin square]{The $\boldsymbol{GF(2)^{\otimes 2}}$ 
geometry of the Mermin square}
\label{twiceGF} 

The remarkable algebraic geometrical properties of
a system of two spin-$\frac{1}{2}$ particles discussed in the
previous section can be given a more appropriate setting if we
employ the concept of projective geometry over rings, in
particular that of projective lines def\/ined over f\/inite rings. The
most prominent, and at f\/irst sight counterintuitive, feature of
ring geometries (of dimension two and higher) is the fact that two
distinct points/lines need not to have a unique connecting
line/meeting point \cite{SanigaDouble,Veldkamp}. As a
result, such geometries feature new concepts like neighbour and
distant, which turns out to be relevant for our geometrical
interpretation of mutual unbiasedness, complementarity and
non-locality in quantum physics. All these features are intimately
connected with the structure of the set of zero divisors of the
ring. As this kind of geometry has until recently been virtually
unknown to the physics community, we shall start from scratch and
recollect f\/irst some basic def\/initions, concepts and properties of
rings (see, e.g., \cite{fr,mcd,ra}).

A {\it ring} is a set $R$ (or, more specif\/ically, ($R,+,\times$))
with two binary operations, usually called addition ($+$) and
multiplication ($\times$), such that $R$ is an Abelian group under
addition and a~semigroup under multiplication, with multiplication
being both left and right distributive over addition\footnote{It
is customary to denote multiplication in  a ring simply by
juxtaposition, using $ab$ in place of $a \times b$, and we shall
follow this convention here.}. A ring in which the multiplication
is commutative is a commutative ring. A ring $R$ with a
multiplicative identity 1 such that 1$r$ = $r$1 = $r$ for all $r
\in R$ is a ring with unity. A ring containing a f\/inite number of
elements is a f\/inite ring. In what follows the word ring will
always mean a commutative ring with unity. An element $r$ of the
ring $R$ is a {\it unit} (or an invertible element) if there
exists an element $r^{-1}$ such that $rr^{-1} = r^{-1} r=1$. The
element $r^{-1}$, uniquely determined by $r$,  is called the
multiplicative inverse of $r$. The set of units forms a group
under multiplication. A (non-zero) element $r$ of $R$ is said to
be a (non-trivial) {\it zero-divisor} if there exists $s \neq 0$
such that $sr= rs=0$. An element of a f\/inite ring is either a unit
or a zero-divisor. A ring in which every non-zero element is a
unit is a {\it field}; f\/inite (or Galois) f\/ields, often denoted by
$GF(q$), have $q$ elements and exist only for $q = p^{n}$, where
$p$ is a prime number and $n$ a positive integer. The smallest
positive integer $s$ such that $s1=0$, where $s1$ stands for $1 +
1 + 1 + \cdots + 1$ ($s$ times), is called the {\it
characteristic} of $R$; if $s1$ is never zero, $R$ is said to be
of characteristic zero. An {\it ideal} ${\cal I}$ of $R$ is a
subgroup of $(R,+)$ such that $a{\cal I} = {\cal I}a \subset {\cal
I}$ for all $a \in R$. An ideal of the ring $R$ which is not
contained in any other ideal but $R$ itself is called a {\it
maximal}  ideal. If an ideal is of the form $Ra$ for some element
$a$ of $R$ it is called a {\it principal} ideal, usually denoted
by~$\langle a \rangle$. A ring with a unique maximal ideal is a
{\it local} ring. Let $R$ be a ring and ${\cal I}$ one of its
ideals. Then $\overline{R} \equiv R/{\cal I} = \{a + {\cal I} ~|~
a \in R\}$ together with addition $(a + {\cal I}) + (b + {\cal I})
= a + b +  {\cal I}$ and multiplication $(a + {\cal I})(b + {\cal
I}) = ab +  {\cal I}$ is a ring, called the quotient (or factor)
ring of $R$ with respect to ${\cal I}$; if ${\cal I}$ is maximal,
then $\overline{R}$ is a f\/ield. A very important ideal of a ring
is that one represented by the intersection of all maximal ideals;
this ideal is called the {\it Jacobson radical}. Finally, we
mention a couple of relevant examples of rings: a polynomial ring,
$R[x]$, viz. the set of all polynomials in one variable $x$ and
with coef\/f\/icients in the ring $R$, and the ring $R_{\otimes}$ that
is a~(f\/inite) direct product of rings, $R_{\otimes} \equiv R_{1}
\otimes R_{2} \otimes \cdots \otimes R_{n}$, where both addition
and multiplication are carried out componentwise and where the
individual rings need not be the same.

Given a ring $R$  with unity, the general linear group of
invertible 2$\times$2 matrices with entries in~$R$, a pair
($\alpha, \beta$) $\in R^{2}$ is called {\it admissible} over $R$
if there exist $\gamma, \delta \in R$ such that \cite{her}
\begin{gather}\label{eq7}
\left(
\begin{array}{cc}
\alpha & \beta \\
\gamma & \delta \\
\end{array}
\right) \in {\rm GL}(2,R).
\end{gather}
The projective line over $R$, denoted as $PR(1)$, is def\/ined as
the set of classes of ordered pairs $(\varrho \alpha, \varrho
\beta)$, where $\varrho$ is a unit and $(\alpha, \beta)$ is
admissible \cite{Veldkamp,her,bh1,bh2,hav}. Such a
line carries two non-trivial, mutually complementary relations of
neighbour and distant. In particular, its two distinct points
$X$:=$(\varrho \alpha, \varrho \beta)$ and $Y$:=$(\varrho \gamma,
\varrho \delta)$ are called {\it neighbour} (or, {\it
parallel}) if
\begin{gather}\label{eq8}
\left(
\begin{array}{cc}
\alpha & \beta \\
\gamma & \delta \\
\end{array}
\right) \notin {\rm GL}(2,R)
\end{gather}
and {\it distant} otherwise, i.e., if condition \eqref{eq7} is met. The
neighbour relation is ref\/lexive (every point is obviously
neighbour to itself) and symmetric (i.e., if $X$ is neighbour to
$Y$ then $Y$ is neighbour to $X$ too), but, in general, not
transitive (i.e., $X$  being neighbour to $Y$ and $Y$ being
neighbour to $Z$ does not necessarily mean that $X$ is neighbour
to $Z$ -- see, e.g., \cite{Veldkamp,her,hav}). Given
a point of $PR(1)$, the set of all neighbour points to it will be
called its {\it neighbourhood}\footnote{To avoid any confusion,
the reader should be cautious that some authors (e.g.
\cite{bh1,hav}) use this term for the set of {\it distant}
points instead.}.  For a
{\it finite commutative} ring $R$, equation~\eqref{eq7} reads
\begin{gather}
 \det \left(
\begin{array}{cc}
\alpha & \beta \\
\gamma & \delta \\
\end{array}
\right) \in R^{*},\label{eq9}
\end{gather}
and equation~\eqref{eq8} reduces to
\begin{gather}
\det \left(
\begin{array}{cc}
\alpha & \beta \\
\gamma & \delta \\
\end{array}
\right) \in R \backslash R^{*},\label{eq10}
\end{gather}
where $R^{*}$ denotes the set of {\it units} (invertible elements)
and $R \backslash R^{*}$ stands for the set of {\it zero-divisors}
of $R$ (including the trivial zero divisor 0). Obviously, if $R$ is a f\/ield then `neighbour'
simply reduces to `identical' and `distant' to `dif\/ferent'; for given the fact that 0 is the only
zero-divisor in a f\/ield, equation~\eqref{eq10} reduces to $\alpha \delta - \gamma \beta = 0$, which indeed
implies that $(\gamma, \delta) = (\varrho \alpha, \varrho \beta)$. 

\begin{table}[t]\small
\caption{Addition and multiplication in 
 $R_{\perp}$ ({\it top}) and in $GF(4) \simeq GF(2)[x]/\langle x^{2} + x + 1 \rangle$ ({\it bottom}).}
 
 \vspace{-3mm}
 
\begin{center}
\begin{tabular}{||c|cccc||}
\hline \hline
$+$ & $0$ & $1$ & $x$ & $x+1$ \\
\hline
$0$ &   $0$ &   $1$ &   $x$ & $x+1$  \\
$1$ &   $1$ &   $0$ &   $x+1$ & $x$  \\
$x$ &   $x$ &   $x+1$ & $0$ & $1$ \\
$x+1$ & $x+1$ & $x$ &   $1$ & $0$ \\
\hline \hline
\end{tabular}~~~~~~~~
\begin{tabular}{||c|cccc||}
\hline \hline
$\times$ & $0$ & $1$ & ~~$x$~~ & $x+1$  \\
\hline
$0$ & $0$ & $0$ & $0$ & $0$  \\
$1$ & $0$ & $1$ & $x$ & $x+1$  \\
$x$ & $0$ & $x$ & $x$ & $0$  \\
$x+1$ & $0$ & $x+1$ & $0$ & $x+1$  \\
\hline\hline
\end{tabular}
\end{center}

\begin{center}
\begin{tabular}{||c|cccc||}
\hline \hline
$+$ & $0$ & $1$ & $x$ & $x+1$ \\
\hline
$0$ & $0$ & $1$ & $x$ & $x+1$  \\
$1$ & $1$ & $0$ & $x+1$ & $x$  \\
$x$ & $x$ & $x+1$ & $0$ & $1$ \\
$x+1$ & $x+1$ & $x$ & $1$ & $0$ \\
\hline \hline
\end{tabular}~~~~~~~
\begin{tabular}{||c|cccc||}
\hline \hline
$\times$ & $0$ & $1$ & $x$ & $x+1$  \\
\hline
$0$ & $0$ & $0$ & $0$ & $0$  \\
$1$ & $0$ & $1$ & $x$ & $x+1$  \\
$x$ & $0$ & $x$ & $x+1$ & $1$  \\
$x+1$ & $0$ & $x+1$ & $1$ & $x$  \\
\hline \hline
\end{tabular}
\label{table4}
\end{center}\vspace{-3mm}
\end{table}

\begin{figure}[t]
\centerline{\includegraphics[width=5.5cm]{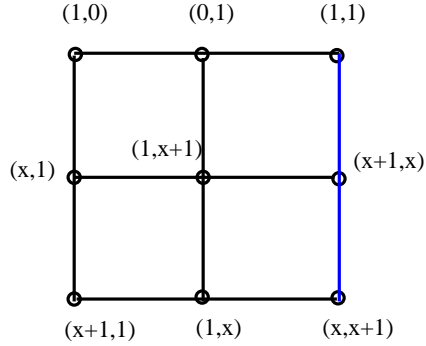}}
\caption{An illustration of the structure of the projective line
over $R_{\perp}$. If two distinct points are joined by a line,
they are distant; if not, they are neighbour.} \label{figure4}
\end{figure}

To illustrate the concept, and to meet the f\/irst relevant example
of ring geometry in quantum theory, we shall consider the
projective line def\/ined over the ring of Galois double numbers
$R_{\perp} \equiv GF(2)^{\otimes 2}$~\cite{SanigaPlanat1}. The
ring $R_{\perp}$ is of characteristic two and consists of the four
elements $0$, $1$, $x$, $x+1$ subject to the addition and multiplication
rules given in Table~\ref{table4},
as it can readily be verif\/ied from its isomorphism to the quotient
ring $GF(2)[x]/\langle x^2-x \rangle$ \cite{SanigaTriple1}; 
it is important to realize at this point that although the addition table
of $R_{\perp}$ is the same as that of the corresponding 
Galois f\/ield, $GF(4) \cong GF(2)[x]/\langle x^{2} + x + 1 \rangle$ (compare the left-hand
sides of Table~\ref{table4}), the two rings substantialy dif\/fer in their 
multiplication tables (the right-hand sides of Table~\ref{table4}), because  $GF(4)$,
like any f\/ield, does not possess any non-trivial zero-divisors.
As `1'
is the only unit of~$R_{\perp}$, from equation~\eqref{eq9} we f\/ind that the
associated projective line, $PR_{\perp}(1)$, features altogether
nine points out of which (i) seven points are represented by pairs
where at least one entry is a unit, namely
\begin{gather*}
(1,0), \ (1,x), \ (1,x+1), \ (1,1),\nonumber \\
(0,1), \ (x,1), \ (x+1,1),
\end{gather*}
and (ii) two points have both coordinate entries zero-divisors, not of the same ideal, viz.
\begin{gather*}
(x,x+1), \ (x+1,x).
\end{gather*}
To reveal the f\/ine structure of the line we pick up 
three distinguished points $U$:=$(1,0)$, $V$:=$(0,1)$ 
and $W$:=$(1,1)$, representing nothing but the ordinary 
projective line of order two embedded in $PR_{\perp}(1)$, which are obviously pairwise distant
and whose neighbourhoods are readily found to read
\begin{gather*}
U:\ \  (1,x), \ (1,x+1), \ (x,x+1), \ (x+1,x), \nonumber \\
V:\ \ (x,1), \ (x+1,1), \ (x,x+1), \ (x+1,x), \nonumber \\
W:\ \ (1,x), \ (1,x+1), \ (x,1), \ (x+1,1).
\end{gather*}
Now, as the coordinate system on this line can always be chosen in such a way that the coordinates
of any three mutually distant points are made identical to those of $U$, $V$ and $W$, from
the last three expressions we discern that
 the neighbourhood of any point of the line features four distinct points, 
 the neighbourhood of any two distant points have two points in common 
 (which makes the neighbour relation non-transitive) and the 
 neighbourhoods of any three mutually distant points have no element in common. 
 The nine points of the line $PR_{\perp}(1)$ can thus be arranged into a 
 $3 \times 3$ array as shown in Fig.~\ref{figure4} (see also Fig.~2 of 
 Ref.~\cite{SanigaPlanat2} for a dif\/ferent, ``conic'' representation of $PR_{\perp}$).
This array has an important property that all the points in the
same row and/or column are pairwise distant. Moreover, a closer
look at Fig.~\ref{figure3} reveals that one triple of mutually
distant points, that located in the third (blue) column, dif\/fers
from all others in having both coordinates of all the three points
of the same character, namely either zero-divisors (the points
$(x,x+1)$ and $(x+1,x))$, or units (the point $(1,1)$). After
identifying, in an obvious way, the observables of  the f\/irst (or
the second) Mermin square in \eqref{Mermin} with the points of
$PR_{\perp}(1)$, one immediately sees that the concept {\it
mutually commuting} translates ring geometrically into {\it
mutually distant} and that the ``Bell-borne'' specif\/ic
character of the observab\-les of the third column has its
geometrical counterpart in the above-mentioned distinguishing
properties of the coordinates of the corresponding points.

\subsection[The three pencil configurations and the projective line over $GF(2)^{\otimes 3}$]{The 
three pencil conf\/igurations and the projective line over $\boldsymbol{GF(2)^{\otimes 3}}$}
\label{threetimesGF}

\begin{table}[t]\small
\caption{Addition and multiplication in  $R_{\triangle}$.}

\vspace{-3mm}

\begin{center}
\begin{tabular}{||c||cccc|cccc||}
\hline \hline
$+$ & $0$ & $1$ & $b$ & $y$ & $r$ & $c$ & $g$ & $m$ \\
\hline\hline
$0$ &      $0$ & $1$ & $b$ & $y$ & $r$ & $c$ & $g$ & $m$  \\
$1$ &      $1$ & $0$ & $y$ & $b$ & $c$ & $r$ & $m$ & $g$ \\
$b$ &      $b$ & $y$ & $0$ & $1$ & $m$ & $g$ & $c$ & $r$ \\
$y$ &      $y$ & $b$ & $1$ & $0$ & $g$ & $m$ & $r$ & $c$ \\
\hline
$r$ &      $r$ & $c$ & $m$ & $g$ & $0$ & $1$ & $y$ & $b$ \\
$c$ &      $c$ & $r$ & $g$ & $m$ & $1$ & $0$ & $b$ & $y$ \\
$g$ &      $g$ & $m$ & $c$ & $r$ & $y$ & $b$ & $0$ & $1$ \\
$m$ &      $m$ & $g$ & $r$ & $c$ & $b$ & $y$ & $1$ & $0$ \\
\hline \hline
\end{tabular}~~~~~~
\begin{tabular}{||l||cccc|cccc||}
\hline \hline
$\times$ & $0$ & $1$ & $b$ & $y$ & $r$ & $c$ & $g$ & $m$ \\
\hline\hline
$0$ &       $0$ & $0$ & $0$ & $0$ & $0$ & $0$ & $0$ & $0$  \\
$1$ &       $0$ & $1$ & $b$ & $y$ & $r$ & $c$ & $g$ & $m$ \\
$b$ &       $0$ & $b$ & $b$ & $0$ & $0$ & $b$ & $0$ & $b$ \\
$y$ &       $0$ & $y$ & $0$ & $y$ & $r$ & $g$ & $g$ & $r$ \\
\hline
$r$ &       $0$ & $r$ & $0$ & $r$ & $r$ & $0$ & $0$ & $r$ \\
$c$ &       $0$ & $c$ & $b$ & $g$ & $0$ & $c$ & $g$ & $b$ \\
$g$ &       $0$ & $g$ & $0$ & $g$ & $0$ & $g$ & $g$ & $0$ \\
$m$ &       $0$ & $m$ & $b$ & $r$ & $r$ & $b$ & $0$ & $m$ \\
\hline\hline
\end{tabular}
\label{table5}
\end{center}\vspace{-3mm}
\end{table}

The above-established mutually commuting -- mutually distant
analogy can be extended to a more general ring geometrical
setting, that of the projective line def\/ined over $R_{\triangle}
\equiv GF(2)^{\otimes 3}$~\cite{SanigaPlanat2}. The ring
$R_{\triangle}$, of characteristic two and cardinality eight,
comprises the unity $[1,1,1]\equiv 1$, the trivial zero-divisor
$[0,0,0]\equiv 0$, and six other zero-divisors forming three
pairs, namely $[1,0,0]\equiv b$ and $[0,1,1]\equiv y$,
$[0,1,0]\equiv r$ and $[1,0,1]\equiv c$, $[0,0,1]\equiv g$ and
$[1,1,0]\equiv m$; the entries in each pair are complementary in
the sense that they sum to the unity. The elements of the ring are
subject to the addition and multiplication properties as shown in
Table~\ref{table5}, from where
it follows that the ring has three maximal -- and principal as
well -- ideals
\begin{gather*}
\langle y \rangle=\{0,r,g,y\},\quad \langle c\rangle=\{0,b,g,c\}\quad \mbox{and}\quad \langle m\rangle=\{0,b,r,m\},
\end{gather*}
and three other principal ideals
\begin{gather*}
\langle b\rangle=\{0,b\}=\langle c\rangle\cap \langle m\rangle, \quad
\langle r\rangle=\{0,r\}=\langle y\rangle\cap \langle m\rangle\quad
\mbox{and}\quad \langle g\rangle=\{0,g\}=\langle y\rangle\cap \langle c\rangle.
\end{gather*}
By making use of these facts, the associated projective line,
$PR_{\triangle}(1)$, is found to consist of the following
twenty-seven points~\cite{SanigaPlanat2}: (i) the three
distinguished points (the ``nucleus''),
\begin{gather*}
(1,0),\qquad (0,1),\quad \mbox{and}\quad (1,1),
\end{gather*}
which represent the ordinary projective line over $GF(2)$ embedded in $PR_{\triangle}(1)$;
(ii) six pairs of points of the ``inner shell'' whose coordinates feature both the unity and a zero-divisor,
\begin{gather*}
(1,b),(b,1);\quad (1,y),(y,1);\quad \ldots;
\end{gather*}
and
(iii) six pairs of points of the ``outer shell'' whose coordinates have zero-divisors in both the entries,
\begin{gather*}
(b,y),(y,b);\quad \ldots;\quad (c,y),(y,c);\quad \ldots;
\end{gather*}
which were split into two groups according to as both entries are
composite zero-divisors or not.

The f\/ine structure of this line has thoroughly been investigated
in \cite{SanigaPlanat2} and the most relevant results are here
reproduced in Tables~\ref{table6} to~\ref{table9}, using the notation of Ref.~\cite{SanigaPlanat2}. After
identifying the points of the line with the observables as shown
in Figs.~\ref{figure1} and~\ref{figure2}, from Tables~\ref{table1} and~\ref{table2} we f\/ind out that the
subsets in question provide a perfect match for the geometry of
all the three pencils. The picture, however, is not complete as
one readily realizes when trying, under the given correspondence,
to reproduce Tables~\ref{table2} and~\ref{table3}; in the former case we see that the
two pictures dif\/fer in four places (Table~\ref{table8}), whilst in the latter
case in as many as fourteen entries (Table~\ref{table9})!

\begin{table}[t]\small
\caption{A subset of $PR_{\triangle}(1)$ whose distant/neighbour ($+/-$) 
relations exactly reproduce the commutation relations embodied in the ``entangled'' 
pencil of lines shown in Fig.~\ref{figure1}.}

\vspace{-3mm}

\begin{center}
\begin{tabular}{||l||ccc|cccc||}
\hline\hline
 & $(1,0)$ & $(0,1)$ & $(1,1)$ & $(c,r)$ & $(b,y)$ & $(y,b)$ & $(r,c)$  \\
\hline\hline
$(1,0)$&  $-$ & $+$ & $+$ & $-$ & $-$ & $-$ & $-$  \\
$(0,1)$&  $+$ & $-$ & $+$ & $-$ & $-$ & $-$ & $-$  \\
$(1,1)$&  $+$ & $+$ & $-$ & $+$ & $+$ & $+$ & $+$  \\
\hline
$(c,r)$&  $-$ & $-$ & $+$ & $-$ & $-$ & $-$ & $+$  \\
$(b,y)$&  $-$ & $-$ & $+$ & $-$ & $-$ & $+$ & $-$  \\
$(y,b)$&  $-$ & $-$ & $+$ & $-$ & $+$ & $-$ & $-$  \\
$(r,c)$&  $-$ & $-$ & $+$ & $+$ & $-$ & $-$ & $-$  \\
\hline\hline
\end{tabular}
\label{table6}
\end{center}\vspace{-3mm}
\end{table}

\begin{table}[t]\small
\caption{A subset of $PR_{\triangle}(1)$ whose
distant/neighbour relations match the commutation relations
exhibited by the ``unentangled'' pencil of lines shown in
Fig.~\ref{figure2}, {\it left}; the identical table is obtained
if we exchange the order of coordinates, which f\/its the geometry
of the pencil of Fig.~\ref{figure2}, {\it right}.}

\vspace{-3mm}

\begin{center}
\begin{tabular}{||l||ccc|cccc||}
\hline\hline
 & $(1,0)$ & $(0,1)$ & $(1,1)$ & $(y,1)$ & $(b,1)$ & $(c,1)$ & $(r,1)$  \\
\hline\hline
$(1,0)$&  $-$ & $+$ & $+$ & $+$ & $+$ & $+$ & $+$  \\
$(0,1)$&  $+$ & $-$ & $+$ & $-$ & $-$ & $-$ & $-$  \\
$(1,1)$&  $+$ & $+$ & $-$ & $-$ & $-$ & $-$ & $-$  \\
\hline
$(y,1)$&  $+$ & $-$ & $-$ & $-$ & $+$ & $-$ & $-$  \\
$(b,1)$&  $+$ & $-$ & $-$ & $+$ & $-$ & $-$ & $-$  \\
$(c,1)$&  $+$ & $-$ & $-$ & $-$ & $-$ & $-$ & $+$  \\
$(r,1)$&  $+$ & $-$ & $-$ & $-$ & $-$ & $+$ & $-$  \\
\hline\hline
\end{tabular}\label{tabel7}
\end{center}\vspace{-3mm}
\end{table}

\begin{table}[t]\small
\caption{A subset of $PR_{\triangle}(1)$ that is
the best match for the geometry of the observables given in Table~\ref{table2}; 
the two conf\/igurations dif\/fer in four places indicated by an
exclamation mark.}

\vspace{-3mm}

\begin{center}
\begin{tabular}{||l||cccc|cccc||}
\hline\hline
 & $(y,1)$ & $(b,1)$ & $(r,1)$ & $(c,1)$ & $(1,c)$ & $(1,b)$ & $(1,r)$ & $(1,y)$  \\
\hline\hline
$(y,1)$&  $-$ & $+$ & $-$ & $-$ & $-$! & $+$ & $-$ & $-$\\
$(b,1)$&  $+$ & $-$ & $-$ & $-$ & $-$ & $-$ & $+$ & $+$\\
$(r,1)$&  $-$ & $-$ & $-$ & $+$ & $+$ & $+$ & $-$ & $-$\\
$(c,1)$&  $-$ & $-$ & $+$ & $-$ & $-$ & $-$ & $+$ & $-!$\\
\hline
$(1,c)$&  $-!$ & $-$ & $+$ & $-$ & $-$ & $-$ & $+$ & $-$\\
$(1,b)$&  $+$ & $-$ & $+$ & $-$ & $-$ & $-$ & $-$ & $+$\\
$(1,r)$&  $-$ & $+$ & $-$ & $+$ & $+$ & $-$ & $-$ & $-$\\
$(1,y)$&  $-$ & $+$ & $-$ & $-!$ & $-$ & $+$ & $-$ & $-$\\
\hline\hline
\end{tabular}
\label{table8} 
\end{center}\vspace{-3mm}
\end{table}

\begin{table}[t]\small
\caption{A subset of $PR_{\triangle}(1)$ that is
the best match for the geometry of the observables given in Table~\ref{table3}; 
the two conf\/igurations dif\/fer in fourteen places indicated by
an exclamation mark.}

\vspace{-3mm}

\begin{center}\begin{tabular}{||l||ccc|cccc||}
\hline\hline
  & $(1,0)$ & $(0,1)$ & $(1,1)$ & $(c,r)$ & $(b,y)$ & $(y,b)$ & $(r,c)$  \\
\hline\hline
$(y,1)$&  $+$ & $-$ & $-$ & $+$ & $+$ & $-$ & $-$\\
$(b,1)$&  $+$ & $-$ & $-$ & $-$ & $-$ & $+$ & $-!$\\
$(r,1)$&  $+$ & $-$ & $-$ & $+!$ & $-$ & $-!$ & $-!$\\
$(c,1)$&  $+$ & $-$ & $-$ & $-!$ & $-!$ & $+!$ & $+!$\\
\hline
$(1,c)$&  $-$ & $+$ & $-$ & $+$ & $+!$ & $-!$ & $-$\\
$(1,b)$&  $-$ & $+$ & $-$ & $-$ & $+$ & $-$ & $-!$\\
$(1,r)$&  $-$ & $+$ & $-$ & $-$ & $-!$ & $-$ & $+$\\
$(1,y)$&  $-$ & $+$ & $-$ & $-!$ & $-$ & $+$ & $+!$\\
\hline\hline
\end{tabular}
\label{table9} 
\end{center}\vspace{-3mm}
\end{table}

\begin{figure}[t]
\centerline{\includegraphics[width=5cm]{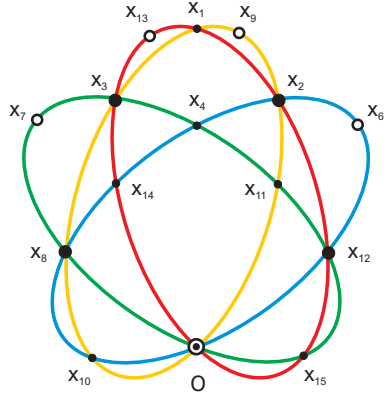}}
\caption{The structure and mutual relation between the four
maximal ideals, represented by points of four distinct ellipses,
of $R_{\diamond}$.} \label{figure21}
\end{figure}

\begin{figure}[t]
\centerline{\includegraphics[width=7cm]{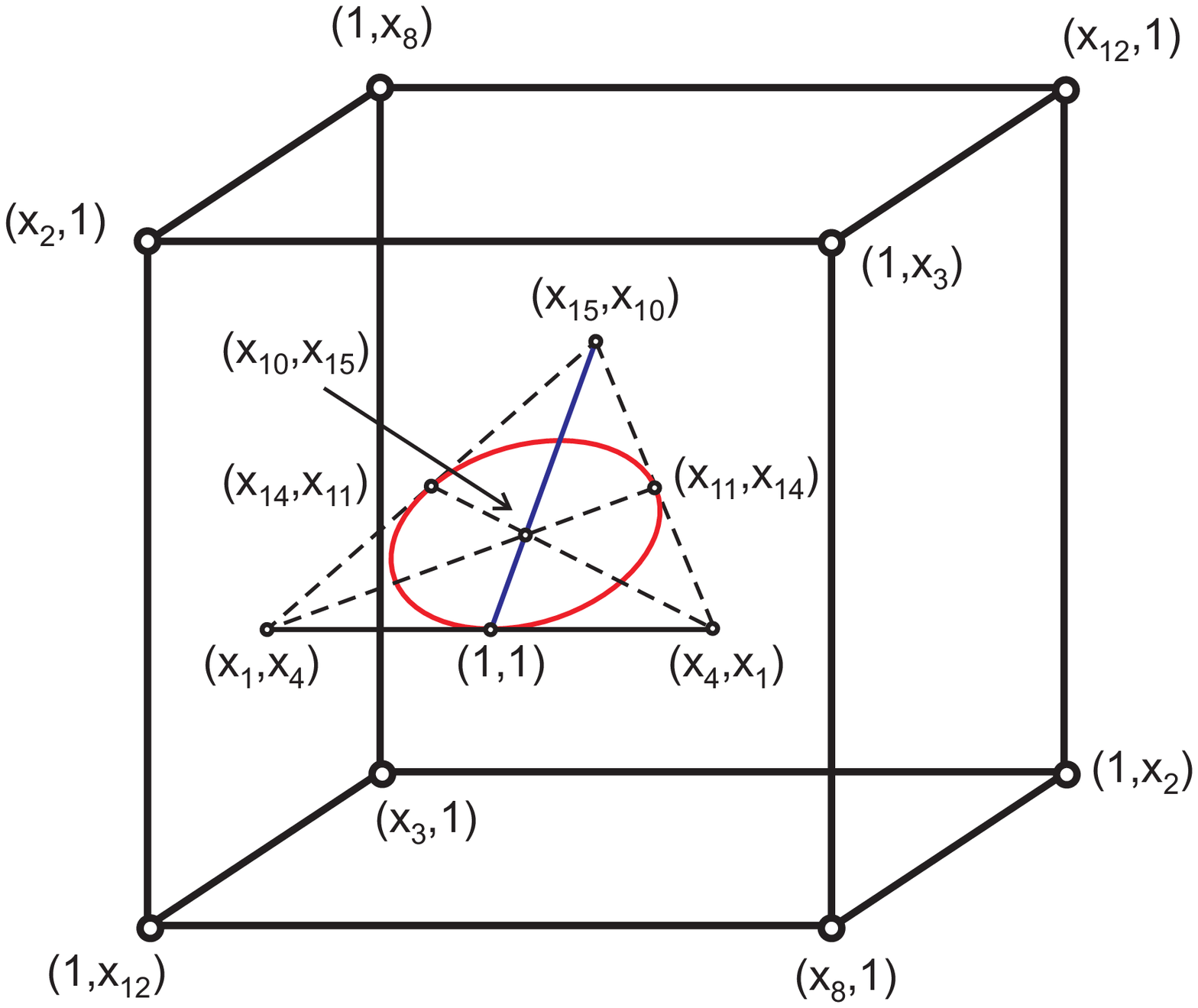}}
\caption{Two distinct subsets  of $PR_{\diamondsuit}(1)$ 
reproducing the structure of both the kernel and the shell of the full conf\/iguration
of observables characterizing two-qubit systems. Unfortunately, this framework is insuf\/f\/icient to harbour the
coupling between the two objects (compare with Fig.~\ref{figure3}).}
\label{figure22}
\end{figure}

\subsection[Towards a fuller picture: the projective line over $GF(2)^{\otimes 4}$]{Towards a 
fuller picture: the projective line over $\boldsymbol{GF(2)^{\otimes 4}}$}
\label{fourtimesGF} 

These last observations clearly indicate that
higher order rings  have to be employed to obtain a satisfactory
picture of the behaviour of two spin-$\frac{1}{2}$ particles. As
an important intermediate step to reach this goal seems to be the
structure of the projective line def\/ined over $R_{\diamondsuit}
\equiv GF(2)^{\otimes 4}$.

We will not go into much detail here and simply observe that the
16 elements of $R_{\diamondsuit}$ can be represented in the
following form
\begin{gather*}
 x_0=[0,0] \equiv 0,\quad x_1=[0,1],\quad x_2=[0,a],\quad x_3=[0,b],\nonumber \\
 x_4=[1,0],\quad x_5=[1,1] \equiv 1,\quad x_6=[1,a],\quad x_7=[1,b],\nonumber \\
 x_8=[a,0],\quad x_9=[a,1],\quad x_{10}=[a,a],\quad x_{11}=[a,b],\nonumber \\
 x_{12}=[b,0],\quad x_{13}=[b,1],\quad x_{14}=[b,a],\quad x_{15}=[b,b].
\end{gather*}
which stems from the fact that $R_{\diamondsuit} \cong
R_{\perp}\otimes R_{\perp}$ and from the representation of
$R_{\perp}$ given in Section~\ref{twiceGF} after identifying $a=x$ and
$b=x+1$. The f\/ifteen zero-divisors of $R_{\diamondsuit}$ form four
maximal ideals, whose composition and mutual relation are depicted
in Fig.~\ref{figure21}. Although yielding the trivial Jacobson radical
($\{x_{0}\}$), any triple of them share one more element, and
there are altogether four such distinguished elements: $x_{2}$,
$x_{3}$, $x_{8}$ and $x_{12}$. The associated projective line,
$PR_{\diamondsuit}(1)$, is easily found to contain subsets whose
properties reproduce properly not only those of Mermin's squares
(like $PR_{\perp}(1)$) and of the three pencil-borne geometries
(like $PR_{\triangle}(1)$), but also a subset which accounts for
the behaviour of the observables forming the ``outer'' shell (the
cube) in Fig.~\ref{figure3}. This particular subset consists of  eight points
whose coordinates feature the unity and one of the above-mentioned
distinguished zero-divisors, as illustrated in Fig.~\ref{figure22}.  So, the
structure of $PR_{\diamondsuit}(1)$  is a proper ring geometrical
setting for the observables of both the ``inner'' and ``outer''
shells when considered separately. Yet, it fails to provide a
correct picture for the coupling between the two shells, because
it implies that no observable from one shell commutes with any
observable from the other one, which is clearly not the case.  To
glue the two pictures thus clearly necessitates to look at
projective lines over higher order, and possibly non-commutative
rings and/or allied algebras.

\section{Conclusion}
The f\/ifteen observables/operators characterizing the interaction
of two spin-$\frac{1}{2}$ particles were found to exhibit two
distinct, yet intimately connected, algebraic geometrical
structures, consi\-de\-red f\/irst as points of the ordinary projective
plane of order two and then as points of projective lines def\/ined
over $GF(2)^{\otimes n}$, with $n = 2,3$ and 4. In the f\/irst
picture, the observables are regarded as three pencils of lines.
These pencil-conf\/igurations, each featuring seven points, share a
line, and a line in any of them comprises three observables. All
the lines in each pencil carry mutually commuting operators; in
one of the pencils, which we call the kernel, the observables on
two lines share a base of Bell states. The three
operators on any line in each pencil represent a row or column of
some Mermin's  ``magic'' square. An inherent geometrical
nature of Mermin's squares is shown to be captured by the
structure of the projective line def\/ined over  $GF(2)^{\otimes
2}$, that of all the three pencils, when taken together, by the
line over $GF(2)^{\otimes 3}$, whereas the behavior of the kernel
and its complement (the cube-shell), when considered separately,
is reproduced by the properties of the line over $GF(2)^{\otimes
4}$. To complete the picture, it just remains to f\/ind a ring line,
or a very similar object, that would also account for the coupling
between the kernel and its complement.

To close this paper, a group-theoretical comment is in order. For
$N$ qubits, the Lie group $U(2^{N})$ and the chain $U(2^{N})
\supset SU(2^{N})$ play an important role. According to a theorem
credited to Racah \cite{kib1}, for a semi-simple Lie group $G$ of
order $r$ and rank $l$, a complete set of $\frac{1}{2}(r+l)$
commuting operators can be constructed in the enveloping algebra
of $G$. For $N=2$ qubits ($d=2^{2}$), the relevant group is
$SU(4)$ for which $r=d^{2}-1=15$ and $l=d-1=3$. In this case, we
have $\frac{r}{l}=d+1=5$ MUBs corresponding to a set of $l=d-1=3$
commuting operators taken from a complete, with respect to
$SU(4)$, set of $\frac{1}{2}(r+l)=\frac{1}{2}(d-1)(d+2)=9$
operators. These matters are presently the object of our
investigations (see also \cite{kib2}).

\subsection*{Acknowledgements}

This work was partially supported by the
Science and Technology Assistance Agency under the contract $\#$
APVT--51--012704, the VEGA project $\#$ 2/6070/26 (both from
Slovak Republic) and by the trans-national ECO-NET project $\#$
12651NJ ``Geometries Over Finite Rings and the Properties of
Mutually Unbiased Bases'' (France). We are grateful to Dr. Petr
Pracna for a~number of fruitful comments/remarks and for creating
the last two f\/igures. One of the authors (M.S.) would like to
thank the warm hospitality extended to him by the Institut
FEMTO-ST in Besan\c con and the Institut de Physique Nucl\'{e}aire in
Lyon.

\LastPageEnding

\end{document}